\documentclass[aps,prd,onecolumn,nofootinbib,preprint]{revtex4-1}
		\usepackage{hyperref}
                       \usepackage{graphicx}				% Use pdf, png, jpg, or eps§ with pdflatex; use eps in DVI mode
                       \usepackage{amssymb}
                       \usepackage{amsfonts,amsmath,amssymb,graphics,epsfig}
                       \usepackage{comment}
                       
\usepackage{footnote}

\begin{document}
  
                       \title{Caustic Formation upon Shift Symmetry Breaking}
                       \author{Klaountia Pasmatsiou}
                       \affiliation{\\ {CERCA, Department of Physics, Case Western Reserve University}\\ {\small 10900 Euclid Avenue, Cleveland, Ohio 44106, USA}}

                       \email{kxp265@case.edu}
                      % \date{\today}						% Activate to display a given date or no date
                       
                       \begin{abstract}
                       We examine how the breaking of shift symmetry affects the formation of caustics for the standard canonical kinetic theory as well as for the DBI theory. We show in this case, that the standard canonical kinetic theory is caustic free but the same does not always apply for the DBI model. We make similar arguments about the conformal Galileons. Finally, it is shown that the simple wave condition is not invariant under field redefinitions and the meaning of simple waves after symmetry breaking is discussed. 
                       \end{abstract}
                       
                       \maketitle
                       
\section{Introduction}

The formation of caustics indicates the breakdown of a low-energy effective field theory (EFT) since the second derivative of the field blows up. The method of characteristics is a useful tool in solving partial differential equations (PDEs) and fully describes the propagation of a signal given a background. Along a characteristic, a PDE transforms into an ordinary differential equation allowing us to identify the existence of caustics for any scalar field model.  In previous studies, the focus was on generic scalar field models that enjoy a global shift-symmetry \cite{Babichev:2016hys,Mukohyama:2016ipl}. As was pointed out in \cite{Babichev:2016hys}, a generic wave solution in k-essence models assuming planar symmetry does not keep its form while propagating, leading to the formation of caustics unlike in the canonical case where caustics are not present. In particular, for the canonical case, the characteristics are always parallel to each other so intersections do not occur. In k-essence, however, the characteristics are no longer parallel to each other, leading to intersections and thus the formation of caustics. It is worth pointing out that an interesting link between k-essence and pressureless perfect fluid has been established in \cite{Babichev:2017lrx} where a caustic-free completion is achieved through a complex scalar field.

Later, it was shown that the canonical case is not the only caustic-free model. The Dirac-Born-Infield (DBI) model is also caustic free when simple wave solutions in a Minkowski background are considered \cite{Mukohyama:2016ipl}. Work has also been done to go beyond the planar symmetry when studying the generalized Galileon Lagrangians where a link has been suggested between the absence of caustics and the existence of a global symmetry for spherical waves when the shift-symmetry is preserved \cite{deRham:2016ged}.   

The motivation behind this work is to present a possible relation between symmetries and caustics in the context of scalar field theories. Many scalar field models have been proposed over the years to explain the late-time accelerated expansion of the Universe. Although these models, exhibit interesting cosmological features, they may lead to the formation of caustics. The presence of caustics is very unappealing and without a possible UV completion of these models, they indicate the breakdown of particular EFTs. Caustics have also been studied in a series of modified theories of gravity like Horava-Lifshitz gravity\cite{Mukohyama:2009tp,Setare:2010gy} or TeVes \cite{Contaldi:2008iw}. However the effects of gravity are beyond the scope of this work and we restrict ourselves to scalar fields that live on flat spacetime.

The manuscript is organized as follows: In section \ref{DOF} we provide the results of applying the method of characteristics to models that break the shift symmetry by introducing an explicit $\phi$ dependence in the Lagrangian. We show that by taking the appropriate limit, we recover the results shown in \cite{Babichev:2016hys,Mukohyama:2016ipl}. Furthermore, we apply the method of characteristics to the $\phi$ dependent standard kinetic term and trivially show that the model is free of caustics. In section \ref{DBI}, we show that breaking the shift symmetry in the DBI model likely leads to the formation of caustics unlike in the shift symmetric case. The result applies to the conformal generalized DBI-Galileon operators also. Finally, in section \ref{genwaves} we discuss what simple waves mean upon shift symmetry breaking and we show that the simple wave condition is not invariant under field redefinitions. After discussing our results in section \ref{Discussion} we provide the reader with details of the method of characteristics in Appendix A.

\section{Dependence on the field}\label{DOF}

To begin, consider the k-essence Lagrangian which is an arbitrary function of the standard kinetic term and the field, thus explicitly breaking the shift symmetry, \footnote{ For the rest of the analysis we work in units where $M_{\mathrm{pl}}=1$, the metric is flat $\eta^{\mu \nu}=(-1,1,1,1)$ and we go beyond the shift symmetry ($\phi\rightarrow \phi + c$) by keeping the explicit $\phi$ dependence of the Lagrangian.}
\begin{equation}
L=L(\phi,X),
\end{equation}
where $X\equiv\partial_\mu \phi \partial^\mu \phi$. The equation of motion for the k-essence Lagrangian with respect to the field is found to be
\begin{eqnarray}\label{50}
(2 L_{XX} \dot{\phi}^2-L_X) \ddot{\phi}- 4 L_{XX} {\phi'} \dot{\phi} \dot{\phi'} + (2 L_{XX} {\phi'}^2 + L_X){\phi''}-\frac{1}{2}L_\phi+ X L_{X \phi}=0,
\end{eqnarray}
where $L_{X}\equiv \frac{\partial L}{\partial X}$, $L_{XX}\equiv \frac{\partial^2 L}{\partial X^2}$, $L_{\phi}\equiv \frac{\partial L}{\partial \phi}$ and  $L_{X\phi}\equiv \frac{\partial^2 L}{\partial X \partial \phi}$. From the method of characteristics (see Appendix A for a discussion), we identify the coefficients of the second derivative terms, $A$, $B$, $C$, and the left over piece, $D$, as
\begin{align}
A&=2 L_{XX}\dot{\phi}^2 -L_X,&B&=-2  L_{XX} \phi' \dot{\phi},&C&= 2 L_{XX} {\phi'}^2 +L_X, &D&= -\frac{1}{2}L_\phi +  X L_{X \phi}.
\end{align}
The characteristic curves describe the propagation of a signal. Their form is determined by the characteristic equation which gives the following two sets of solutions
\begin{eqnarray}\label{xia}
\xi_\pm=\frac{-2 \dot{\phi} \phi' L_{XX} \pm \sqrt{L_{X}^2 + 2 X L_{X} L_{XX}}}{-L_X + 2 \dot{\phi}^2 L_{XX}}.
\end{eqnarray}
For later convenience, we define the following quantities
\begin{align}\label{cs}
u&\equiv-\frac{\phi'}{\dot{\phi}},&c_s^2&\equiv\left(1+2X \frac{L_{XX}}{L_{X}}\right)^{-1},
 \end{align}  
\begin{equation}\label{lkh}
\xi_{\pm}\equiv\frac{u \pm c_s}{1\pm u c_s},
\end{equation}  
where $\xi$ is the wavefront velocity, i.e. the speed which is relevant for causality \cite{Shore:2007um,Babichev:2007dw}. In the high frequency and momentum limit, the wave front velocity is the velocity of short wavelength perturbations of the field on a Minkowski background and in this particular limit is equivalent to the phase velocity. Physically, the speed of sound is the speed of propagation of information while $u$ is the velocity of the k-essence field. Using the compatibility relation (\ref{A9}), we can bring it to the following form which is more conveniently written in terms of the previous definitions
\begin{equation} \label{he}
\left(\frac{1}{c_s} \frac{dX}{d\sigma_{\mp}} \frac{1}{X}\mp\frac{2}{1-u^2} \frac{du}{d\sigma_{\mp}}\right) \frac{d\sigma_\mp}{dt}= \frac{2 A_4 \tau  (1\pm u c_s) }{c_s X}.
\end{equation} 
where $A_4=\frac{D}{A}$ and $\sigma_{\pm}$ parametrize the characteristic curves. These equations along with
\begin{align}
\frac{dx}{d\sigma_{+}}&= \xi_{+} \frac{dt}{d\sigma_{+}}&\frac{dx}{d\sigma_{-}}&= \xi_{-} \frac{dt}{d\sigma_{-}},
\end{align}  
give us the set of the coupled differential equations to be solved for $\dot\phi$, $\phi'$, $x$ and $t$. This will enable us to determine the existence of caustics in the models under study.

\subsection{Special cases}

To reproduce the results of \cite{Mukohyama:2016ipl,Babichev:2016hys}, let us examine the case where only second derivate terms appear in the EOM ($D=0$) . For the moment, we ignore the $\phi$ dependence and integration of (\ref{he}) gives us
\begin{equation} \label{101}
\int \frac{1}{X c_s} dX \mp \ln\!\left(\frac{1+u}{1-u}\right)= \Gamma (\sigma_\pm),
\end{equation} 
where $\Gamma (\sigma_\pm)$ are called Riemann invariants \cite{supersonic}.  For simple waves, those that satisfy the relation
\begin{eqnarray} \label{has}
\ddot\phi \phi'' - \dot \phi'^2=0,
\end{eqnarray}
one of the Riemann invariants needs to be a constant. If neither of the Riemann invariants are a constant, then we are dealing with general wave solutions. 
 
\subsubsection{Standard canonical kinetic term}
For $L=-\frac{1}{2}X$ we have $\xi_\pm=\pm 1$ and the solutions are simply
\begin{align}\label{12}
\dot{\phi}&=\phi' + C(\sigma_-), &\dot{\phi}&=-\phi' + C(\sigma_+).
\end{align}
Thus, the characteristics are straight lines in phase space as well as in the $x$-$t$ plane (since $\xi$ is a constant). The characteristics that belong in the same family, will not intersect leaving the model free of caustics. In other words, the family of curves with constant $\sigma_{+}$ or $\sigma_{-}$ are straight lines, parallel to each other.  Further, for this case we have
\begin{equation}
\xi=v_{ph}|_{k \rightarrow \infty}=v_g=c_s.
 \end{equation} 

\subsubsection{DBI kinetic term}

Similarly, for the DBI Lagrangian $L= -\sqrt{1+X}$, the characteristic curves are
\begin{eqnarray}  \label{xi}
\xi_{\pm} = \frac{1}{1+\phi'^2} (-\dot{\phi} \phi' \pm \sqrt{1 + X}).
\end{eqnarray}
At first glance, this appears to be worrying since the characteristic curves depend on the field. However, we observe that  $\xi_{\pm}$ can be constant only if $\dot \phi$ is linear in $\phi'$,\footnote{This is equivalent to the completely exceptional condition as pointed out in \cite{Mukohyama:2016ipl,Deser:1998wv}.}
\begin{eqnarray} \label {100}
\dot{\phi} = \mp a \phi' + b,
\end{eqnarray}
where $a$, $b$ are constants and $ b^2=1-a^2$. We can easily verify that (\ref{100}) is the general solution of  (\ref{101}).  This leads to the result $\xi_{\pm}=\pm a$. With this in mind, we can express $X$ in terms of $a$ and $b$, 
\begin{eqnarray} 
X= -\dot{\phi}^2 + \phi'^2 = (1-a^2) \phi'^2 \pm 2 a b \phi' - b^2.
\end{eqnarray}
Also, the square of the speed of sound is linear in $X$. Using (\ref{lkh}), we derive
\begin{eqnarray} 
c_s= \sqrt{1+\alpha X},
\end{eqnarray}
where $\alpha \equiv \frac{1-a^2}{b^2}$. Without loss of generality, we may set $\alpha=1$ and, as a consequence, we obtain $1-a^2=b^2$. Note that the speed of sound can be either subluminal or superluminal. If we choose  $\alpha>0$,  then for a timelike field, $X<0 $, the propagation is subluminal, while  for a spacelike field, $X>0$, the propagation is superluminal. Alternatively, for $\alpha<0$, these cases are swapped. Finally, for the DBI model, we have 
\begin{equation}
\xi=v_{ph}|_{k \rightarrow \infty}=v_g=\frac{u\pm c_s}{1 \pm u c_s}.
 \end{equation} 
As was pointed out in \cite{deRham:2016ged}, the reason that the DBI model might have such a special place among $P(X)$ theories is that in addition to the shift symmetry, it enjoys a global symmetry that is related to invariance under higher-dimensional rotations and boosts. This might protect the theory against caustics and as it was shown, the same applies for the pure Galileon and the DBI Galileon which enjoy a global nonrelativistic and relativistic Galilean symmetry, respectively. The question we are going to explore is what happens when we break the shift symmetry in these models.

\subsection{Breaking the shift symmetry in the standard canonical kinetic term}
A simple way to break shift symmetry is to introduce a scaling function. For a general  $L=-\frac{1}{2}f(\phi) X$, the equation of motion is\footnote{Note that the negative signs are kept in this equation to be consistent with (\ref{50}).}
\begin{equation} \label{haha}
-\Box \phi - \frac{1}{2 f} \frac{df}{d\phi} X=0.
\end{equation}
The solution to the characteristic equation (\ref{xia}) remains
\begin{equation}\label{hoha}
\xi_{\pm}=\left(\frac{dx}{dt}\right)_{\pm}=\frac{1}{A}\left(B \pm \sqrt{B^2 - A C}\right)= \pm 1,
\end{equation}
while the compatibility relation (\ref{he})  leads to
\begin{equation} \label{ss}
\frac{1}{c_s}  \frac{dX}{X}\mp\frac{2 du}{1-u^2} = \frac{2}{c_s} \left(- \frac{1}{2 f} df \left(\frac{dt}{d\phi}\right)_\mp\right)\dot{\phi}  (1\pm u c_s) ,
\end{equation} 
where 
\begin{align}
\left(\frac{d \phi}{dt}\right)_{\pm}=\dot{\phi}\left(\frac{1-u^2}{1\pm u c_s}\right).
\end{align}
Finally from (\ref{cs}), since $L_{XX}=0$,we see that $c_s=1$, so (\ref{ss}) simplifies to
\begin{equation}
\frac{dX}{X}\mp\frac{2 du}{1-u^2}  =  - \frac{1}{ f(\phi)} df(\phi).
\end{equation} 
This can be easily integrated to find
\begin{eqnarray} \label{ba}
\dot{\phi}=\pm \phi' + \frac{C(\sigma_\mp)}{\sqrt{f(\phi)}}.
\end{eqnarray}
Since $\xi\pm$ are constant, it is clear that the characteristic curves in the $x$-$t$ plane are parallel straight lines so they do not intersect. This means that a model with any scaling function multiplying the standard kinetic term does not lead to the formation of caustics. Naturally, this had to occur since by a field redefinition we can always remap the starting Lagrangian to the standard form, $L=-\frac{1}{2}X$, reestablishing the shift symmetry.

It is worth pointing out that since in the analysis without the $\phi$ dependence, the solutions are simple waves (\ref{has}), they satisfy
 \begin{align} \label{b}
\ddot{\phi}|_{\sigma_\pm}&= \frac{d\tau}{d\sigma_\mp} \frac{d\sigma_\mp}{dt} & \phi''|_{\sigma_\pm}&= \frac{d \chi}{d\sigma_\mp} \frac{d\sigma_\mp}{dx} & \dot{\phi'}|_{\sigma_\pm}= \frac{d \chi}{d\sigma_\mp} \frac{d\sigma_\mp}{dt}=\frac{d \tau}{d\sigma_\mp} \frac{d\sigma_\mp}{dx}.
\end{align}
We can easily check that solutions of the form (\ref{ba}) do not satisfy the simple wave condition. We see that the characteristics in the $\dot{\phi}-\phi'$  plane are not straight lines; however, by a field redefinition, they can be transformed back to straight lines. We will return to this point in section \ref{genwaves}.

Another way of breaking the shift symmetry is by introducing a potential as
\begin{equation}\label{dir}
L=-\frac{1}{2}f(\phi) X-V(\phi).
\end{equation}
Caustics are not generated for this case either. The reason is that $\xi_{\pm}$,  which is associated with the kinetic structure of the theory, is still given by (\ref{hoha}) and thus is constant. 

\subsection{Comments on Conformal Galileons}

In general, the conformal Galileons are obtained by taking the nonrelativistic limit of the conformal DBI Galileon (flat brane in AdS space). The conformal Galileons preserve the shift symmetry; however, they break the flat Galilean symmetry \cite{Nicolis:2008in,deRham:2010eu}. As an example, we consider the Lagrangian for the quadratic conformal Galileon:
\begin{equation}\label{con}
L_2=-\frac{1}{2} e^{- 2 \pi} \partial _\mu \pi \partial^\mu \pi.
\end{equation}
As stated above, a simple scaling function multiplying the standard kinetic term can be redefined back to $L=-\frac{1}{2}X$. Hence, by performing the field redefinition $\phi=e^{-\pi}$ on (\ref{con}), we arrive at the standard canonical kinetic term with wave equation $\Box \phi= 0$. We, thus, see that the quadratic conformal Galileon is caustic free.

The next step is to add the cubic conformal Galileon,
\begin{equation}\label{27}
L_3=-\frac{1}{2} \partial_\mu \pi \partial^\mu \pi \Box \pi - \frac{1}{4} (\partial_\mu \pi)^4,
\end{equation}
and study the characteristics of $L_2 + a_3 L_3$. By imposing the simple wave condition $(\ddot{\pi} \pi'' -\dot{\pi'}^2=0)$, the equation of motion is linear in second derivatives, allowing application of the method of characteristics. The characteristic curves are
\begin{equation}
\xi_{\pm}= \frac{2 a_3 e^{2 \pi} \pi' \dot{\pi} \pm \sqrt{1+ 4 a_3 e^{2 \pi} (-\dot{\pi}^2 +\pi'^2)+ 3 {a_3}^2 e^{4 \pi} (-\dot{\pi}^2 +\pi'^2)^2}}{1- a_3 e^{2 \pi} (3 \dot{\pi}^2 -\pi'^2)}.
\end{equation}
This expression for $\xi_\pm$ is very similar to the characteristic curves studied in \cite{Babichev:2016hys} for the $L=X+{X^2}/2$ case\footnote{This Lagrangian is written in the notation of \cite{Babichev:2016hys} which used the $(+,-,-,-)$ signature and defined $X=\frac{1}{2} \partial_\mu \phi \partial^\mu \phi$.}, which is known to contain caustics. This is not surprising since the pure Galileon part of (\ref{27}) vanishes for simple waves in $1+1$.  This leads to the conclusion that when imposing the simple wave condition, the combination of $L_2+ a_3 L_3$ is not caustic free. It is expected that the inclusion of the quartic and quintic conformal Galileons will exhibit caustics as well since the explicit $\phi$ dependence in combination with the complicated form of the higher-derivative operators will lead to $\phi$-dependent wavefront velocities as we will see later for the conformal DBI case.

\section{The DBI model with $\phi$ dependence}\label{DBI}

Next, we consider the introduction of a scaling function in the DBI model. Given the Lagrangian,
\begin{eqnarray} \label{hehehe}
L= -f(\phi) \sqrt{1+X} + f(\phi),
\end{eqnarray}
the EOM has the following form:
\begin{eqnarray} \label{120}
\ddot {\phi} (1+ \phi'^2) + \phi'' (-1 + \dot{\phi}^2) - 2 \dot{\phi'} \dot{\phi} \phi' + \frac{1}{f(\phi)} \frac{d f(\phi)}{d\phi}\left((1+X)^{\frac{3}{2}} - (1+X)\right)=0.
\end{eqnarray}
Since the coefficients of the second derivative terms are independent of the scaling function, the solution to the characteristic equation is the same as in the $\phi$-independent case (\ref{xi}). There, we were able to prove that $\dot{\phi}$ is linear in $\phi'$ and $\xi_\pm=\mp a$. In this case, we have to solve the compatibility relation which includes a $\phi$-dependent piece,
\begin{equation}\label{ll}
\frac{1}{c_s}  \frac{dX}{X}\mp\frac{2 du}{1-u^2} =\frac{2}{f(\phi)} d f(\phi)  \frac{ c_s^2 -c_s }{X}.
\end{equation} 
By looking at (\ref{xi}), we see that  that the wavefront velocity can no longer be constant since the linear relation between $\dot\phi$ and $\phi'$ with constant coefficients (\ref{100})  does not satisfy the equation of motion due to the source term. A first attempt at preserving a linear relationship is to include $\phi$ dependence in the coefficients, $\dot{\phi} = \mp a(\phi) \phi' + b(\phi)$. This, however, fails since plugging the ansatz into the EOM (\ref{120}) results in a differential equation for the coefficients with explicit $\phi'$ (or $\dot{\phi}$) dependence, inconsistent with the initial choice. In fact, this is not surprising. We know for the standard DBI model that $\xi_{\pm}$ is a constant only when the linear relationship with constant coefficients is a valid solution. Since $\xi_{\pm}$ is unchanged with the introduction of a scaling function, it cannot be constant in this case. With $\xi_{\pm}$ not constant it is more difficult to determine if caustics exist. However, having $\xi_\pm$ as a function of $\dot\phi$ and $\phi'$  is very likely to lead to intersections of the characteristics. For simple wave solutions, the characteristics have to be parallel straight lines in the $x$-$t$ plane. For general waves, the characteristics are not restricted to be straight lines and whether they intersect or not is less clear. It is worth pointing out that the existence of caustics might not be fatal to the theory unless they arise for very generic initial/boundary conditions.

We conclude that introducing a scaling function, which breaks the shift symmetry, into the DBI model is very likely to introduce caustics which once again might indicate a close connection between global symmetries and freedom from caustics. The result does not come as a surprise since a similar result has been pointed out in related cases \cite{Felder:2002sv,Mukohyama:2002vq,Goswami:2010rs,Barnaby:2004nk}. For example, caustics have been found for specific cosmological setups where generic classes of potentials, $f(\phi)$, have been considered. In contrast to this work, where we are applying the method of characteristics to the most general case, the main focus there was on homogeneous scalar fields as well as inhomogeneous scalar fields with small gradients. It is worth pointing out that just as with the standard kinetic term, through the use of a field redefinition, we can find cases where caustics will not appear in the DBI model. As an example consider a Lagrangian of the form $L=-\sqrt{1+f(\phi)X}$. For this case, caustics clearly do not form as we can see immediately from the fact that a simple field redefinition can bring the result back to the shift symmetric DBI model. 

The next step is to break the shift symmetry for the DBI model by adding a potential
\begin{eqnarray} \label{heheha}
L= -\sqrt{1+X} - V(\phi).
\end{eqnarray}
The structure in this case is different; nevertheless, this is likely to introduce caustics similar to the inclusion of a scaling function. Just as in that case, a linear relation between $\dot\phi$ and $\phi'$ with constant coefficients (\ref{100})  does not satisfy the equation of motion thus leading to non-constant $\xi$.

\subsection{Comment on the conformal DBI-Galileon}

Similar to the comments on the conformal Galileon, we can consider the quadratic conformal DBI-Galileon Lagrangian \cite{deRham:2010eu}, given by 
\begin{eqnarray} 
L= \lambda' e^{-4 \lambda \pi} (-\sqrt{1+e^{2 \lambda \pi} X} + 1).
\end{eqnarray}
Here, by performing the field redefinition $\phi = \frac{1}{\lambda} (1-e^{\lambda \pi})$, the Lagrangian is brought to the form of (\ref{hehehe}) with $f(\phi)= (1- \lambda \phi)^{-4}$,
\begin{eqnarray}
L= -(1- \lambda \phi)^{-4} \sqrt{1+X} + (1- \lambda \phi)^{-4}.
\end{eqnarray}
Thus we clearly see that the conformal DBI-Galileon is likely to contain caustics.

\section{Simple waves when breaking the shift symmetry}\label{genwaves}

In this section, we discuss what simple waves mean when we break the shift symmetry. We show that the simple wave condition is not invariant under field redefinitions. As an example, consider the transformation
\begin{eqnarray} \label{hehhe}
\phi \rightarrow f(\psi).
\end{eqnarray}
The simple wave condition becomes
\begin{eqnarray} \label{heha}
\ddot{\phi} \phi''-\dot{\phi'}^2= f' f'' (\ddot{\psi} {\psi'}^2 + \dot{\psi}^2 {\psi''} - 2 \dot{\psi} \psi' \dot{\psi'})  + f'^2 (\ddot{\psi} \psi'' - \dot{\psi'}^2).
\end{eqnarray}
The first thing to notice when looking at (\ref{heha}) is that if $\phi$ is a simple wave solution then the left-hand side of the equation is zero. However, due to the leftover piece $ f' f'' (\ddot{\psi} {\psi'}^2 + \dot{\psi}^2 {\psi''} - 2 \dot{\psi} \psi' \dot{\psi'})$, the simple wave condition in $\phi$ is not the same with the simple wave condition in $\psi$. We recover simple waves for $\psi$ only for the trivial case of $f \rightarrow\psi$. Let us make this point more clear by doing a simple example. Consider the quadratic + cubic Galileon, $L_2+L_3= -\frac{1}{2} \partial_{\mu } f(\psi)\partial^{\mu } f(\psi)-\frac{1}{2}\partial_{\mu } f(\psi)\partial^{\mu } f(\psi) \Box f({\psi})$, where the $\phi \rightarrow f(\psi)$ transformation was performed. The EOM in $1+1$ is 
\begin{eqnarray} \label{hehahh}
\Box\psi-\frac{2}{f'} \left[ f' f'' (\ddot{\psi} {\psi'}^2 + \dot{\psi}^2 {\psi''} - 2 \dot{\psi} \psi' \dot{\psi'})  + f'^2 (\ddot{\psi} \psi'' - \dot{\psi'}^2)\right]+\frac{f''}{f'} X=0,
\end{eqnarray}
where $X \equiv \partial_\mu \psi \partial^\mu \psi$. Notice that the term in the square brackets is precisely the field redefined simple wave condition (\ref{heha}). If we naively set $\ddot{\psi} \psi'' - \dot{\psi'}^2=0$, then the characteristic equation would depend on $\dot\psi$ and $\psi'$, making it unclear if the general solution of the remainder has caustics. To have a clear picture, we should set the term in the square brackets equal to zero  and find the solution to the remaining part using the compatibility relation. This reduces to a form very close to (\ref{haha}). In this case we find, $\dot{\psi}=\pm\psi' + \frac{C(\sigma_{\mp})}{f'}$ which does not satisfy $f'^2 (\ddot{\psi} \psi'' - \dot{\psi'}^2)=0$, it does however satisfy $f' f'' (\ddot{\psi} {\psi'}^2 + \dot{\psi}^2 {\psi''} - 2 \dot{\psi} \psi' \dot{\psi'})  + f'^2 (\ddot{\psi} \psi'' - \dot{\psi'}^2)=0$.  For this type of wave, caustics do not form as we see from (\ref{hoha}). 

On the other hand, by keeping $f' f'' (\ddot{\psi} {\psi'}^2 + \dot{\psi}^2 {\psi''} - 2 \dot{\psi} \psi' \dot{\psi'})  + f'^2 (\ddot{\psi} \psi'' - \dot{\psi'}^2)$ in the EOM, and linearizing in $\psi$, the explicit $\psi$ dependence leads to $\psi$-dependent wavefronts. Again this is very likely to lead to caustics, unless highly tuned initial conditions are constructed to avoid them. 

In contrast, it is useful to look at a case in which we are not able to perform a field redefinition to simple waves. Consider the following example where the shift symmetry, as well as the Galilean symmetry, are broken,
\begin{eqnarray} \label{heh}
L_2+L_3=-\frac{1}{2}f(\psi) X -\frac{1}{2} f(\psi) X \Box \psi.
\end{eqnarray}
The EOM reads
\begin{eqnarray} \label{hehe}
 \Box \psi -\frac{2}{f}\left[ f (\ddot \psi \psi'' - \dot{\psi'}^2)+ f' (\psi'' \psi'^2+\ddot\psi  \dot\psi^2- 2 \dot{\psi} \psi' \dot{\psi'})\right]+ \frac{1}{2} \frac{f'}{ f} X  - \frac{1}{2} \frac{f''}{ f} X^2=0.
\end{eqnarray}
First, let us do the naive thing as before and set $f(\ddot \psi \psi'' - \dot{\psi'}^2)=0$. We immediately see that we run into problems as in the previous example. Setting this condition to zero might be inconsistent, as simple waves might not be solutions to this model. Alternatively, we could again try to set the term in the square brackets to zero and look for a solution to the remainder, $ \Box \psi + \frac{1}{2} \frac{f'}{ f} X-\frac{1}{2} \frac{f''}{ f}  X^2=0$, which is caustic free. An explicit calculation, though, shows that the compatibility relation cannot be solved analytically. That means that we cannot check whether the solution of the remainder satisfies the constraint from the term in the square brackets. This is in contrast to the previous case, where setting the constraint to zero is exactly the simple wave condition in the $\phi$ field. However, it would be interesting to see if one could find solutions to $ \Box \psi + \frac{1}{2} \frac{f'}{ f} X-\frac{1}{2} \frac{f''}{ f}  X^2=0$ which also satisfy the term in the square brackets being zero. That would impose a new constraint, and hence this would be a special class of waves that satisfies this condition and is caustic free.

\section{Discussion}\label{Discussion}

In this work, we explored the formation of caustics for the standard canonical kinetic theory as well the standard DBI model and the conformal Galileons  when breaking the shift symmetry. We showed that multiplying the canonical kinetic term by any function of $\phi$ does not lead to the formation of caustics. This does not come as a surprise since we can always perform a field redefinition of the kinetic term and bring it back to its original form which obeys the shift symmetry. The freedom from caustics remains even with the inclusion of a potential. Additionally, we explored the conformal Galileons, leading to the result that the inclusion of the cubic conformal Galileon is not caustic free (at least for simple waves), since it reduces to a P(X) model which is known to contain caustics. We pointed out that the conformal Galileons preserve the shift symmetry; however, they break the flat Galilean symmetry, hence making the connection of caustics and global symmetries more evident, as has been previously suggested. 

Next we considered breaking the shift symmetry in the DBI. For this case, we discovered that the linear relation between $\dot\phi$ and $\phi'$ no longer holds, hence leaving the wavefront speed to depend on the initial choice of the field which is a function of $x$ and $t$. Unlike the shift symmetric case, this means that the DBI model is not manifestly caustic free when breaking the shift symmetry; however, we do not exclude the possibility that caustic-free solutions can still be constructed. By multiplying the DBI Lagrangian by a function of $\phi$, it is obvious that a field redefinition does not exist to bring it back to the shift symmetric form. In addition, we considered a variant of the DBI model which is field redefinable back to its shift-symmetric form and thus is caustic free.

Finally, we showed what simple waves mean when we break the shift symmetry. We saw that  the simple wave condition is not invariant under field redefinitions; hence, it might be different for different field choices. Thus, we should be careful when imposing the simple wave condition as it may be inconsistent with the EOM and mislead us to believe that caustics might form. 

Any model which is field redefinable to a caustic-free model is naturally free of caustics too. When breaking the global symmetries in our theory though, the models that trivially remain caustic free are the ones that have a wavefront velocity equal to the speed of light. The close connection between global symmetries and caustics is evident throughout this work. We notice that breaking global symmetries in our models results in a difficulty of analytically integrating the compatibility relation. Breaking a global symmetry, also, removes a conserved current. This might suggest a connection between the two; that is, the integrability of the compatibility relation relies on the existence of a global symmetry. More work is required to establish this connection.  
\newline
\noindent \textbf{Acknowledgments:} I would like to thank Claudia de Rham for suggesting this problem and for useful discussions throughout its development. Additionally, I would like to thank Craig Copi for introducing me to the method of characteristics and numerous discussions throughout this work. Finally, I would like to thank Kurt Hinterbichler for useful discussions related to Galileon theories and Glenn Starkman for useful feedback and guidance. K.P. acknowledges support from CWRU.

\appendix\label{fro}
\section{The method of Characteristics}

The method of characteristics applies to second-order partial differential equations that are linear in second derivatives and can take the general form
\begin{equation}
A(\phi,\dot{\phi},\phi') \ddot{\phi}+ 2 B(\phi,\dot{\phi},\phi')  \dot{\phi}' + C(\phi,\dot{\phi},\phi')\phi'' + D(\phi,\dot{\phi},\phi')=0,
\end{equation}
where the dot represents the partial derivative with respect to time and prime with respect to $x$. To find solutions to a hyperbolic PDE, it is sufficient to specify Cauchy boundary conditions, that is $\phi$, $\dot{\phi}$ and $\phi'$ along a curve. Consider such a curve parametrized by $\sigma$ so that $x(\sigma)$ and $t(\sigma)$ along this curve. Then the second derivatives of the field can be calculated as,
\begin{align}
\frac{d \dot{\phi}}{d\sigma}&= \ddot{\phi}\frac{dt}{d\sigma} + \dot{\phi}' \frac{dx}{d\sigma}& \frac{d \phi'}{d\sigma}=&\dot{\phi'}\frac{dt}{d\sigma} +  \phi'' \frac{dx}{d\sigma}.
\end{align}
These equations can be solved everywhere except where
\begin{equation}
\begin{vmatrix}
      A      & 2 B & C  \\ 
      \frac{dt}{d\sigma}       & \frac{dx}{d\sigma} & 0 \\
    0     & \frac{dt}{d\sigma}& \frac{dx}{d\sigma} \\ 
    \end{vmatrix}=0=A (\frac{dx}{d\sigma})^2- 2 B \frac{dt}{d\sigma}\frac{dx}{d\sigma} + C(\frac{dt}{d\sigma})^2=A (\frac{dx}{dt})^2- 2 B \frac{dx}{dt} + C.
  \end{equation}
  The solutions to this quadratic equation are
\begin{equation}
\xi_{\pm}=\left(\frac{dx}{dt}\right)_{\pm}=\frac{1}{A} \left(B \pm \sqrt{B^2 - A C}\right),
\end{equation}
where $\xi_{\pm}$ are the two sets of the characteristic curves. Rewriting the equations for the second derivatives as
\begin{align}
 \ddot{\phi} + \dot{\phi}' \frac{dx}{d t}-\frac{d \dot{\phi}}{d t}=&0, & \dot{\phi'} + \phi'' \frac{dx}{dt}-\frac{d \phi'}{dt}=&0,
\end{align}
we can solve these equations along with the original differential equation for the second derivatives. As an example, we solve for $ \ddot  \phi$ to find
\begin{equation}
   \ddot  \phi= \frac{\begin{vmatrix}
      D      & 2 B & C  \\ 
      - \frac{d  \dot{\phi}}{dt}    & \frac{dx}{dt} & 0 \\
      -\frac{d \phi'}{dt}  & 1&\frac{dx}{dt} \\ 
    \end{vmatrix}}{\begin{vmatrix}
   A   & 2B & C  \\ 
      \frac{dt}{d\sigma}       & \frac{dx}{d\sigma} & 0 \\
    0     & \frac{dt}{d\sigma}& \frac{dx}{d\sigma} \\ 
    \end{vmatrix}}.
\end{equation}  
Along the characteristics, the denominator is zero. To ensure that the second derivative is defined along the characteristic, the numerator must also be zero. Enforcing this, we find
\begin{equation}
D (\xi)^2 + (2 B \xi - C) \frac{d \dot{\phi}}{dt} + C \frac{d  \phi'}{dt} \xi=0,
\end{equation}    
and using
\begin{equation}
A \xi^2 - 2 B \xi  + C =0,
\end{equation}  
we have 
\begin{equation}\label{A9}
A_4 +   \frac{d  \dot{\phi}}{dt}  + \frac{1}{\epsilon_{\pm}}  \frac{d  \phi'}{dt}=0,
\end{equation}  
where
\begin{align}
\frac{1}{\epsilon_{\pm}} &= \frac{C}{\xi_{\pm} A}= \xi_{\mp},&A_4=\frac{D}{A}.
\end{align} 
This is called the compatibility relation satisfied by the derivatives along the characteristic curves. 
\bibliography{Untitledcaustic}

\begin{thebibliography}{17}%
\makeatletter
\providecommand \@ifxundefined [1]{%
 \@ifx{#1\undefined}
}%
\providecommand \@ifnum [1]{%
 \ifnum #1\expandafter \@firstoftwo
 \else \expandafter \@secondoftwo
 \fi
}%
\providecommand \@ifx [1]{%
 \ifx #1\expandafter \@firstoftwo
 \else \expandafter \@secondoftwo
 \fi
}%
\providecommand \natexlab [1]{#1}%
\providecommand \enquote  [1]{``#1''}%
\providecommand \bibnamefont  [1]{#1}%
\providecommand \bibfnamefont [1]{#1}%
\providecommand \citenamefont [1]{#1}%
\providecommand \href@noop [0]{\@secondoftwo}%
\providecommand \href [0]{\begingroup \@sanitize@url \@href}%
\providecommand \@href[1]{\@@startlink{#1}\@@href}%
\providecommand \@@href[1]{\endgroup#1\@@endlink}%
\providecommand \@sanitize@url [0]{\catcode `\\12\catcode `\$12\catcode
  `\&12\catcode `\#12\catcode `\^12\catcode `\_12\catcode `\%12\relax}%
\providecommand \@@startlink[1]{}%
\providecommand \@@endlink[0]{}%
\providecommand \url  [0]{\begingroup\@sanitize@url \@url }%
\providecommand \@url [1]{\endgroup\@href {#1}{\urlprefix }}%
\providecommand \urlprefix  [0]{URL }%
\providecommand \Eprint [0]{\href }%
\providecommand \doibase [0]{http://dx.doi.org/}%
\providecommand \selectlanguage [0]{\@gobble}%
\providecommand \bibinfo  [0]{\@secondoftwo}%
\providecommand \bibfield  [0]{\@secondoftwo}%
\providecommand \translation [1]{[#1]}%
\providecommand \BibitemOpen [0]{}%
\providecommand \bibitemStop [0]{}%
\providecommand \bibitemNoStop [0]{.\EOS\space}%
\providecommand \EOS [0]{\spacefactor3000\relax}%
\providecommand \BibitemShut  [1]{\csname bibitem#1\endcsname}%
\let\auto@bib@innerbib\@empty
%</preamble>
\bibitem [{\citenamefont {Babichev}(2016)}]{Babichev:2016hys}%
  \BibitemOpen
  \bibfield  {author} {\bibinfo {author} {\bibfnamefont {E.}~\bibnamefont
  {Babichev}},\ }\href {\doibase 10.1007/JHEP04(2016)129} {\bibfield  {journal}
  {\bibinfo  {journal} {JHEP}\ }\textbf {\bibinfo {volume} {04}},\ \bibinfo
  {pages} {129} (\bibinfo {year} {2016})},\ \Eprint
  {http://arxiv.org/abs/1602.00735} {arXiv:1602.00735 [hep-th]} \BibitemShut
  {NoStop}%
%%CITATION = ARXIV:1602.00735;%%
\bibitem [{\citenamefont {Mukohyama}\ \emph {et~al.}(2016)\citenamefont
  {Mukohyama}, \citenamefont {Namba},\ and\ \citenamefont
  {Watanabe}}]{Mukohyama:2016ipl}%
  \BibitemOpen
  \bibfield  {author} {\bibinfo {author} {\bibfnamefont {S.}~\bibnamefont
  {Mukohyama}}, \bibinfo {author} {\bibfnamefont {R.}~\bibnamefont {Namba}}, \
  and\ \bibinfo {author} {\bibfnamefont {Y.}~\bibnamefont {Watanabe}},\ }\href
  {\doibase 10.1103/PhysRevD.94.023514} {\bibfield  {journal} {\bibinfo
  {journal} {Phys. Rev.}\ }\textbf {\bibinfo {volume} {D94}},\ \bibinfo {pages}
  {023514} (\bibinfo {year} {2016})},\ \Eprint
  {http://arxiv.org/abs/1605.06418} {arXiv:1605.06418 [hep-th]} \BibitemShut
  {NoStop}%
%%CITATION = ARXIV:1605.06418;%%
\bibitem [{\citenamefont {Babichev}\ and\ \citenamefont
  {Ramazanov}(2017)}]{Babichev:2017lrx}%
  \BibitemOpen
  \bibfield  {author} {\bibinfo {author} {\bibfnamefont {E.}~\bibnamefont
  {Babichev}}\ and\ \bibinfo {author} {\bibfnamefont {S.}~\bibnamefont
  {Ramazanov}},\ }\href {\doibase 10.1007/JHEP08(2017)040} {\bibfield
  {journal} {\bibinfo  {journal} {JHEP}\ }\textbf {\bibinfo {volume} {08}},\
  \bibinfo {pages} {040} (\bibinfo {year} {2017})},\ \Eprint
  {http://arxiv.org/abs/1704.03367} {arXiv:1704.03367 [hep-th]} \BibitemShut
  {NoStop}%
%%CITATION = ARXIV:1704.03367;%%
\bibitem [{\citenamefont {de~Rham}\ and\ \citenamefont
  {Motohashi}(2017)}]{deRham:2016ged}%
  \BibitemOpen
  \bibfield  {author} {\bibinfo {author} {\bibfnamefont {C.}~\bibnamefont
  {de~Rham}}\ and\ \bibinfo {author} {\bibfnamefont {H.}~\bibnamefont
  {Motohashi}},\ }\href {\doibase 10.1103/PhysRevD.95.064008} {\bibfield
  {journal} {\bibinfo  {journal} {Phys. Rev.}\ }\textbf {\bibinfo {volume}
  {D95}},\ \bibinfo {pages} {064008} (\bibinfo {year} {2017})},\ \Eprint
  {http://arxiv.org/abs/1611.05038} {arXiv:1611.05038 [hep-th]} \BibitemShut
  {NoStop}%
%%CITATION = ARXIV:1611.05038;%%
\bibitem [{\citenamefont {Mukohyama}(2009)}]{Mukohyama:2009tp}%
  \BibitemOpen
  \bibfield  {author} {\bibinfo {author} {\bibfnamefont {S.}~\bibnamefont
  {Mukohyama}},\ }\href {\doibase 10.1088/1475-7516/2009/09/005} {\bibfield
  {journal} {\bibinfo  {journal} {JCAP}\ }\textbf {\bibinfo {volume} {0909}},\
  \bibinfo {pages} {005} (\bibinfo {year} {2009})},\ \Eprint
  {http://arxiv.org/abs/0906.5069} {arXiv:0906.5069 [hep-th]} \BibitemShut
  {NoStop}%
%%CITATION = ARXIV:0906.5069;%%
\bibitem [{\citenamefont {Setare}\ and\ \citenamefont
  {Momeni}(2012)}]{Setare:2010gy}%
  \BibitemOpen
  \bibfield  {author} {\bibinfo {author} {\bibfnamefont {M.~R.}\ \bibnamefont
  {Setare}}\ and\ \bibinfo {author} {\bibfnamefont {D.}~\bibnamefont
  {Momeni}},\ }\href {\doibase 10.1007/s10773-011-0894-8} {\bibfield  {journal}
  {\bibinfo  {journal} {Int. J. Theor. Phys.}\ }\textbf {\bibinfo {volume}
  {51}},\ \bibinfo {pages} {198} (\bibinfo {year} {2012})},\ \Eprint
  {http://arxiv.org/abs/1009.0918} {arXiv:1009.0918 [hep-th]} \BibitemShut
  {NoStop}%
%%CITATION = ARXIV:1009.0918;%%
\bibitem [{\citenamefont {Contaldi}\ \emph {et~al.}(2008)\citenamefont
  {Contaldi}, \citenamefont {Wiseman},\ and\ \citenamefont
  {Withers}}]{Contaldi:2008iw}%
  \BibitemOpen
  \bibfield  {author} {\bibinfo {author} {\bibfnamefont {C.~R.}\ \bibnamefont
  {Contaldi}}, \bibinfo {author} {\bibfnamefont {T.}~\bibnamefont {Wiseman}}, \
  and\ \bibinfo {author} {\bibfnamefont {B.}~\bibnamefont {Withers}},\ }\href
  {\doibase 10.1103/PhysRevD.78.044034} {\bibfield  {journal} {\bibinfo
  {journal} {Phys. Rev.}\ }\textbf {\bibinfo {volume} {D78}},\ \bibinfo {pages}
  {044034} (\bibinfo {year} {2008})},\ \Eprint {http://arxiv.org/abs/0802.1215}
  {arXiv:0802.1215 [gr-qc]} \BibitemShut {NoStop}%
%%CITATION = ARXIV:0802.1215;%%
\bibitem [{\citenamefont {Shore}(2007)}]{Shore:2007um}%
  \BibitemOpen
  \bibfield  {author} {\bibinfo {author} {\bibfnamefont {G.~M.}\ \bibnamefont
  {Shore}},\ }\href {\doibase 10.1016/j.nuclphysb.2007.03.034} {\bibfield
  {journal} {\bibinfo  {journal} {Nucl. Phys.}\ }\textbf {\bibinfo {volume}
  {B778}},\ \bibinfo {pages} {219} (\bibinfo {year} {2007})},\ \Eprint
  {http://arxiv.org/abs/hep-th/0701185} {arXiv:hep-th/0701185 [hep-th]}
  \BibitemShut {NoStop}%
%%CITATION = HEP-TH/0701185;%%
\bibitem [{\citenamefont {Babichev}\ \emph {et~al.}(2008)\citenamefont
  {Babichev}, \citenamefont {Mukhanov},\ and\ \citenamefont
  {Vikman}}]{Babichev:2007dw}%
  \BibitemOpen
  \bibfield  {author} {\bibinfo {author} {\bibfnamefont {E.}~\bibnamefont
  {Babichev}}, \bibinfo {author} {\bibfnamefont {V.}~\bibnamefont {Mukhanov}},
  \ and\ \bibinfo {author} {\bibfnamefont {A.}~\bibnamefont {Vikman}},\ }\href
  {\doibase 10.1088/1126-6708/2008/02/101} {\bibfield  {journal} {\bibinfo
  {journal} {JHEP}\ }\textbf {\bibinfo {volume} {02}},\ \bibinfo {pages} {101}
  (\bibinfo {year} {2008})},\ \Eprint {http://arxiv.org/abs/0708.0561}
  {arXiv:0708.0561 [hep-th]} \BibitemShut {NoStop}%
%%CITATION = ARXIV:0708.0561;%%
\bibitem [{\citenamefont {Courant}\ and\ \citenamefont
  {Friedrichs}(1948)}]{supersonic}%
  \BibitemOpen
  \bibfield  {author} {\bibinfo {author} {\bibfnamefont {R.}~\bibnamefont
  {Courant}}\ and\ \bibinfo {author} {\bibfnamefont {K.~O.}\ \bibnamefont
  {Friedrichs}},\ }\href@noop {} {\emph {\bibinfo {title} {Supersonic Flow and
  Shock Waves}}},\ \bibinfo {type} {Tech. Rep.}\ (\bibinfo  {institution}
  {Interscience Publishers, New York},\ \bibinfo {year} {1948})\BibitemShut
  {NoStop}%
\bibitem [{\citenamefont {Deser}\ \emph {et~al.}(1999)\citenamefont {Deser},
  \citenamefont {McCarthy},\ and\ \citenamefont {Sarioglu}}]{Deser:1998wv}%
  \BibitemOpen
  \bibfield  {author} {\bibinfo {author} {\bibfnamefont {S.}~\bibnamefont
  {Deser}}, \bibinfo {author} {\bibfnamefont {J.~G.}\ \bibnamefont {McCarthy}},
  \ and\ \bibinfo {author} {\bibfnamefont {O.}~\bibnamefont {Sarioglu}},\
  }\href {\doibase 10.1088/0264-9381/16/3/015} {\bibfield  {journal} {\bibinfo
  {journal} {Class. Quant. Grav.}\ }\textbf {\bibinfo {volume} {16}},\ \bibinfo
  {pages} {841} (\bibinfo {year} {1999})},\ \Eprint
  {http://arxiv.org/abs/hep-th/9809153} {arXiv:hep-th/9809153 [hep-th]}
  \BibitemShut {NoStop}%
%%CITATION = HEP-TH/9809153;%%
\bibitem [{\citenamefont {Nicolis}\ \emph {et~al.}(2009)\citenamefont
  {Nicolis}, \citenamefont {Rattazzi},\ and\ \citenamefont
  {Trincherini}}]{Nicolis:2008in}%
  \BibitemOpen
  \bibfield  {author} {\bibinfo {author} {\bibfnamefont {A.}~\bibnamefont
  {Nicolis}}, \bibinfo {author} {\bibfnamefont {R.}~\bibnamefont {Rattazzi}}, \
  and\ \bibinfo {author} {\bibfnamefont {E.}~\bibnamefont {Trincherini}},\
  }\href {\doibase 10.1103/PhysRevD.79.064036} {\bibfield  {journal} {\bibinfo
  {journal} {Phys. Rev.}\ }\textbf {\bibinfo {volume} {D79}},\ \bibinfo {pages}
  {064036} (\bibinfo {year} {2009})},\ \Eprint {http://arxiv.org/abs/0811.2197}
  {arXiv:0811.2197 [hep-th]} \BibitemShut {NoStop}%
%%CITATION = ARXIV:0811.2197;%%
\bibitem [{\citenamefont {de~Rham}\ and\ \citenamefont
  {Tolley}(2010)}]{deRham:2010eu}%
  \BibitemOpen
  \bibfield  {author} {\bibinfo {author} {\bibfnamefont {C.}~\bibnamefont
  {de~Rham}}\ and\ \bibinfo {author} {\bibfnamefont {A.~J.}\ \bibnamefont
  {Tolley}},\ }\href {\doibase 10.1088/1475-7516/2010/05/015} {\bibfield
  {journal} {\bibinfo  {journal} {JCAP}\ }\textbf {\bibinfo {volume} {1005}},\
  \bibinfo {pages} {015} (\bibinfo {year} {2010})},\ \Eprint
  {http://arxiv.org/abs/1003.5917} {arXiv:1003.5917 [hep-th]} \BibitemShut
  {NoStop}%
%%CITATION = ARXIV:1003.5917;%%
\bibitem [{\citenamefont {Felder}\ \emph {et~al.}(2002)\citenamefont {Felder},
  \citenamefont {Kofman},\ and\ \citenamefont {Starobinsky}}]{Felder:2002sv}%
  \BibitemOpen
  \bibfield  {author} {\bibinfo {author} {\bibfnamefont {G.~N.}\ \bibnamefont
  {Felder}}, \bibinfo {author} {\bibfnamefont {L.}~\bibnamefont {Kofman}}, \
  and\ \bibinfo {author} {\bibfnamefont {A.}~\bibnamefont {Starobinsky}},\
  }\href {\doibase 10.1088/1126-6708/2002/09/026} {\bibfield  {journal}
  {\bibinfo  {journal} {JHEP}\ }\textbf {\bibinfo {volume} {09}},\ \bibinfo
  {pages} {026} (\bibinfo {year} {2002})},\ \Eprint
  {http://arxiv.org/abs/hep-th/0208019} {arXiv:hep-th/0208019 [hep-th]}
  \BibitemShut {NoStop}%
%%CITATION = HEP-TH/0208019;%%
\bibitem [{\citenamefont {Mukohyama}(2002)}]{Mukohyama:2002vq}%
  \BibitemOpen
  \bibfield  {author} {\bibinfo {author} {\bibfnamefont {S.}~\bibnamefont
  {Mukohyama}},\ }\href {\doibase 10.1103/PhysRevD.66.123512} {\bibfield
  {journal} {\bibinfo  {journal} {Phys. Rev.}\ }\textbf {\bibinfo {volume}
  {D66}},\ \bibinfo {pages} {123512} (\bibinfo {year} {2002})},\ \Eprint
  {http://arxiv.org/abs/hep-th/0208094} {arXiv:hep-th/0208094 [hep-th]}
  \BibitemShut {NoStop}%
%%CITATION = HEP-TH/0208094;%%
\bibitem [{\citenamefont {Goswami}\ \emph {et~al.}(2010)\citenamefont
  {Goswami}, \citenamefont {Nandan},\ and\ \citenamefont
  {Sami}}]{Goswami:2010rs}%
  \BibitemOpen
  \bibfield  {author} {\bibinfo {author} {\bibfnamefont {U.~D.}\ \bibnamefont
  {Goswami}}, \bibinfo {author} {\bibfnamefont {H.}~\bibnamefont {Nandan}}, \
  and\ \bibinfo {author} {\bibfnamefont {M.}~\bibnamefont {Sami}},\ }\href
  {\doibase 10.1103/PhysRevD.82.103530} {\bibfield  {journal} {\bibinfo
  {journal} {Phys. Rev.}\ }\textbf {\bibinfo {volume} {D82}},\ \bibinfo {pages}
  {103530} (\bibinfo {year} {2010})},\ \Eprint {http://arxiv.org/abs/1006.3659}
  {arXiv:1006.3659 [hep-th]} \BibitemShut {NoStop}%
%%CITATION = ARXIV:1006.3659;%%
\bibitem [{\citenamefont {Barnaby}(2004)}]{Barnaby:2004nk}%
  \BibitemOpen
  \bibfield  {author} {\bibinfo {author} {\bibfnamefont {N.}~\bibnamefont
  {Barnaby}},\ }\href {\doibase 10.1088/1126-6708/2004/07/025} {\bibfield
  {journal} {\bibinfo  {journal} {JHEP}\ }\textbf {\bibinfo {volume} {07}},\
  \bibinfo {pages} {025} (\bibinfo {year} {2004})},\ \Eprint
  {http://arxiv.org/abs/hep-th/0406120} {arXiv:hep-th/0406120 [hep-th]}
  \BibitemShut {NoStop}%
%%CITATION = HEP-TH/0406120;%%
\end{thebibliography}%
\end{document}